\documentclass[twocolumn]{autart} 

\usepackage{graphicx} 

\usepackage{amsmath}
\usepackage{amsfonts}
\usepackage{float}
\usepackage{arydshln}

\usepackage{color}

\usepackage{tikz}
\usetikzlibrary{arrows.meta,positioning}

\usepackage{scalerel}
\let\bigopsize\bigoplus
\def\bigominus{{\scalerel*{\boldsymbol\ominus}{\bigopsize}}}
\usepackage{setspace} 

\usepackage{natbib}

\begin{document}

\begin{frontmatter}

\title{Adaptive Learning-based Model Predictive Control for Uncertain Interconnected Systems: A Set Membership Identification Approach \thanksref{footnoteinfo1}} 

\thanks[footnoteinfo1]{Funded by NCCR Automation, the Swiss National Science Foundation (Corresponding Author: Ahmed Aboudonia)
}

\author[Zurich]{Ahmed Aboudonia}\ead{ahmedab@ethz.ch},    
\author[Zurich]{John Lygeros}\ead{lygeros@ethz.ch}               

\address[Zurich]{Automatic Control Laboratory, ETH Zurich}  


\begin{abstract}  
We propose a novel adaptive learning-based model predictive control (MPC) scheme for interconnected systems which can be decomposed into several smaller dynamically coupled subsystems with uncertain coupling. The proposed scheme is mainly divided into two main online phases; a learning phase and an adaptation phase. Set membership identification is used in the learning phase to learn an uncertainty set that contains the coupling strength using online data. In the adaptation phase, rigid tube-based robust MPC is used to compute the optimal predicted states and inputs. Besides computing the optimal trajectories, the MPC ingredients are adapted in the adaptation phase taking the learnt uncertainty set into account. These MPC ingredients include the prestabilizing controller, the rigid tube, the tightened constraints and the terminal ingredients. The recursive feasibility of the proposed scheme as well as the stability of the corresponding closed-loop system are discussed. The developed scheme is compared in simulations to existing schemes including robust, adaptive and learning-based MPC.
\end{abstract}

\end{frontmatter}

\section{Introduction}

Model Predictive Control (MPC) is an advanced control scheme that optimally controls a given system by repeatedly solving an online optimal control problem (OCP) (\cite{kouvaritakis2016model}). MPC was successfully used in various applications such as interconnected systems (e.g. see \cite{muller2017economic} and the references therein).
MPC, however, heavily relies on the availability of an accurate model which is, in many cases, difficult to find and/or complex to use. 
Various efforts were devoted to finding MPC variants that do not require an accurate model. 
These variants include robust, stochastic and learning-based MPC (\cite{saltik2018outlook,mesbah2016stochastic,ren2022tutorial}). 
One drawback of these schemes is that the model is usually not updated online leading to potentially conservative performance. 
Recently, various adaptive and safe learning-based MPC schemes (\cite{hewing2020learning}) have been proposed to ensure safety while enhancing closed-loop performance by improving the model using data collected online. This goal is achieved by combining physics-based knowledge with machine learning and system identification.

In this paper, we propose a novel adaptive learning-based MPC scheme for interconnected systems which can be decomposed into several smaller subsystems. 
Each subsystem is dynamically coupled with a set of other subsystems called the set of neighbours. 
The source of uncertainty in each subsystem is the strength of the coupling with its neighbours.
The proposed scheme learns the uncertainty set online by applying distributed set membership identification on collected data and adapts the MPC ingredients accordingly. 
In particular, the developed control scheme adapts the rigid tube, the prestabilizing controller, the tightened constraint sets and the terminal ingredients by considering them as decision variables in the OCP. 
The developed scheme is compared in simulation to existing schemes including robust MPC, adaptive MPC and learning-based MPC using a network of dynamically coupled double integrators and found to lead to a better trade-off between closed-loop performance and computation cost.
The main contributions and challenges of this work can be listed as follows:
\vspace{-0.25cm}
\begin{itemize}
    \item Unlike many MPC schemes which usually only compute optimal state and input trajectories online, we formulate an MPC problem which can also compute various MPC ingredients in the adaptation phase. The main challenge here is to maintain the convexity of the OCP since these ingredients should generally satisfy non-convex infinite dimensional conditions.
    \item Although set membership identification is extensively used for monolithic systems, we extend this approach here to interconnected systems with uncertain couplings where parameter estimation is not handled by a centralized unit.
    \item We show that the closed-loop system under the developed controller is input-to-state stable. The main challenges here are that the closed-loop system is switched due to updating the MPC ingredients online and that artificial equilibrium points are used. Hence, standard MPC stability arguments do not apply.
    \item We compare the proposed scheme to other off-the-shelf MPC schemes which prove to be successful in many applications.
\end{itemize}
\vspace{-0.25cm}

In Section 2, we formulate the considered control problem. In Section 3, we develop the novel adaptive learning-based MPC scheme. In Section 4, we prove the recursive feasibility of the proposed scheme and the input-to-state stability of the corresponding closed-loop system. In Section 5, we validate the efficacy of the developed scheme by comparing it to other existing schemes in the literature. Finally, we draw some conclusions in Section 6. 
In the sequel, we refer to symmetric, symmetric positive semidefinite and symmetric positive definite matrices of size $n$ as $\mathbb{S}^n$, $\mathbb{S}^n_+$ and $\mathbb{S}^n_{++}$, respectively.
Similarly, we denote diagonal, diagonal positive semidefinite and diagonal positive definite matrices of size $n$ by $\mathbb{D}^n$, $\mathbb{D}^n_+$ and $\mathbb{D}^n_{++}$,  respectively. Additionally, we use $\bigoplus$ and $\bigominus$ to refer to the Minkowski set addition and the Pontryagin set difference, respectively.

\section{Problem Formulation}
 
We consider discrete-time linear time-invariant interconnected systems with a sparse structure.
We assume that such systems can be decomposed into $M$ dynamically-coupled subsystems and that the $i$-th subsystem has a set of in-neighbours $\mathcal{N}_i^-$ including all subsystems which affect its dynamics and a set of out-neighbours $\mathcal{N}_i^+$ including all subsystems whose dynamics are affected by the $i$-th subsystem.
We assume that the dynamics of the $i$-th subsystem is given by,
\vspace{-0.2cm}
\begin{equation}
    \label{sec2_dcn_dyn}
    z_i(t+1) \hspace{-0.1cm} 
    = 
    \hspace{-0.1cm} A_{i}z_i(t)+B_iv_i(t) + E_i \sum\nolimits_{j \in \mathcal{N}_i^-} a_{ij} C_j z_j(t) + d_i(t),
\end{equation}
where $z_i \in \mathbb{R}^{n_i}$, $v_i \in \mathbb{R}^{p_i}$ and $d_i \in \mathbb{R}^{n_i}$ are the state, input and noise vectors of the $i$-th subsystem, $A_{i} \in \mathbb{R}^{n_i \times n_i}$, $B_i \in \mathbb{R}^{n_i \times p_i}$, $C_{i} \in \mathbb{R}^{m_0 \times n_i}$ and $E_i \in \mathbb{R}^{n_i \times m_0}$ are known matrices and $a_{ij} \in \mathbb{R}$ are unknown fixed parameters that represent the coupling strengths. 
We assume that the scalars $a_{ij}$ satisfy $0 \leq a_{ij}^{min} \leq a_{ij} \leq a_{ij}^{max}$ where the bounds $a_{ij}^{min}$ and $a_{ij}^{max}$ are known. 
We also assume that each subsystem is subject to the state and input constraints $z_i(t) \in \mathcal{Z}_i = \{z_i : G_i z_i \leq 1\}$ and $v_i(t) \in \mathcal{V}_i = \{ v_i : H_i v_i \leq 1 \}$ where $G_i \in \mathbb{R}^{q_i \times n_i}$, $H_i \in \mathbb{R}^{r_i \times p_i}$, $\mathcal{Z}_i$ and $\mathcal{V}_i$ are compact and include the origin in their interior. The noise vector $d_i(t)$ is assumed to lie in a known compact polytopic set $\mathcal{D}_i = \{d_i : \Xi_i d_i \leq 1 \}$ that includes the origin in its interior.
Following \cite{riverso2013plug}, we express \eqref{sec2_dcn_dyn} as,
\vspace{-0.2cm}
\begin{equation}
    \label{sec2_dcn_dyn2}
    z_i(t+1) = A_{i}z_i(t)+B_iv_i(t) + E_iw_i(t) + d_i(t),
\end{equation}
\vspace{-0.6cm}
\begin{equation*}
    w_i(t) = \hspace{-0.05cm} \sum\nolimits_{j \in \mathcal{N}_i^-} \hspace{-0.05cm} a_{ij}C_jz_j(t) \in \mathcal{W}_i = \hspace{-0.045cm} \bigoplus\nolimits_{j \in \mathcal{N}_i^-} \hspace{-0.05cm} a_{ij}^{max} C_j Z_j.
\end{equation*}
By denoting the nominal state and input vectors of the $i$-th subsystem by $x_i \in \mathbb{R}^{n_i}$ and $u_i \in \mathbb{R}^{p_i}$, we also define the nominal dynamics of the $i$-th subsystem to be, $x_i(t+1) = A_{i}x_i(t)+B_iu_i(t)$.
Assuming that $v_i(t)=u_i(t)+K_i(z_i(t)-x_i(t))$ where $K_i \in \mathbb{R}^{p_i \times n_i}$ is a stabilizing control gain and defining $e_i(t)=z_i(t)-x_i(t)$, we introduce the error dynamics,
\vspace{-0.2cm}
\begin{equation}
    \label{sec2_dcn_error}
    e_i(t+1) = (A_{i}+B_iK_i)e_i(t)+E_iw_i(t)+d_i(t).
\end{equation}
We consider here a tracking problem where the state is required to follow the target point $x_{r_i}$ which is an equilibrium point of the nominal system. The target point $x_{r_i}$ and the corresponding input $u_{r_i}$ are assumed to satisfy the state and input constraints. For this purpose, we define the artificial equilibrium point $(x_{e_i},u_{e_i})$ satisfying $x_{e_i} = A_{i} x_{e_i} + B_i u_{e_i}$ 
and the cost function
$ J_i = \sum_{k=0}^{N-1} \left( \|x_i(k|t)-x_{e_i}\|_{Q_i}^2 + \|u_i(k|t)-u_{e_i}\|_{R_i}^2 \right) + \|x_{e_i}-x_{r_i}\|_{S_i}^2$
where $x_i(k|t)$ and $u_i(k|t)$ are the $k$-step ahead predicted nominal state and input vectors of the $i$-th subsystem at the $t$-th time instant. 
Note that $(x_{e_i},u_{e_i})$ is used to enlarge the feasible region in tracking problems (\cite{limon2008mpc}) and is considered as an artificial reference that converges asymptotically to the target point.
Although the main aim is to drive the actual state to the target point, we use the nominal state in $J_i$ since we can only predict the behavior of the nominal dynamics.
Note, however, that the closed-loop stability guarantees derived in Section 4 apply to the actual states.
In summary, an MPC problem can be formulated for the $i$-th subsystem with the local OCP,
\vspace{-0.25cm}
\begin{equation}
    \label{sec2_dcn_OCP}
    \begin{aligned}
        \min_{
        \begin{aligned} \end{aligned}
        } 
        J_i \ \ 
        s.t. \left\{
        \begin{aligned}
            & \text{ for all } k \in \{0,...,N-1\}, \\
            & z_i(t) - x_i(0|t) \in \mathcal{Z}_{K_i} , \\
            & x_i(k+1|t)=A_{i}x_i(k|t)+B_iu_i(k|t), \\
            & x_i(k|t) \in \mathcal{X}_{i}, \quad u_i(k|t) \in \mathcal{U}_{i}, \\
            & x_i(N|t) = x_{e_i}=A_{i}x_{e_i}+B_iu_{e_i}, \\
            & x_{e_i} \in \xi \mathcal{X}_{i}, \quad u_{e_i} \in \xi \mathcal{U}_{i},
        \end{aligned}
        \right.
    \end{aligned}
    \vspace{-0.25cm}
\end{equation}
where $x_i(k|t)|_{k \in \{0,\hdots,N}\}, u_i(k|t)|_{k \in \{0,\hdots,N-1\}}, x_{e_i}$ and $u_{e_i}$ are the decision variables, $z_i(t)$ is the measured state of the $i$-th subsystem, $\mathcal{Z}_{K_i}$ is a robust positive invariant (RPI) set for its error dynamics \eqref{sec2_dcn_error} under the disturbances $w_i(t)$ and $d_i(t)$, $\xi \in (0,1)$, $\mathcal{X}_i = \{x_i : G_i x_i \leq g_i\}$ and $\mathcal{U}_i = \{ u_i : H_i x_i \leq h_i\}$ are its tightened constraint sets with $g_i$ and $h_i$ positive scalars. The last three constraints in \eqref{sec2_dcn_OCP} indicate that $x_i(N|t)$ reaches an invariant terminal set satisfying the state and input constraints. 
The scalar $\xi$ is used to ensure that the artificial equilibrium lies in the interior of the constraint sets. This requirement is needed later in the stability proof.
Note that the global prestabilzing control gain is denoted by $K=\operatorname{diag}(K_1,...,K_M)$. Similarly, the global system matrices $A$ and $B$ can be obtained in the obvious way. Following \cite{riverso2013plug}, the control input $v_i(t)=u_i(0|t)+K_i(z_i(t)-x_i(0|t))$ is applied to \eqref{sec2_dcn_dyn} to ensure robust constraint satisfaction.

Interconnected systems with uncertain coupling \eqref{sec2_dcn_dyn} can be exploited in various scenarios with varying-topology networks and large-scale systems. For example, varying-topology networks comprise subsystems which can join and leave the network. While accurate models can be identified offline for the new subsystems, the coupling strength between the network and these subsystems can only be identified online after the plug-in operation. 
Practical applications comprising such systems include microgrids, power systems and vehicle platoons.
Note that considering the whole coupling term as a disturbance might be conservative. This, however, leads to decentralized OCPs rather than distributed ones. Decentralized OCPs are usually more robust against failures and requires less communication and computation. This is more appealing for safety-critical systems with fast dynamics.
Moreover, the inclusion of parametric uncertainties within an additive term might be also conservative. An alternative consideration involves employing adaptive MPC tailored for parametric uncertainties, as applied in monolithic systems (\cite{lorenzen2019robust}). However, the complexity arises in interconnected systems when uncertainties are embedded within coupling terms, necessitating the formulation of a distributed OCP, which imposes substantial communication and computational burdens.
Deceralized OCPs have been widely used with interconnected systems (\cite{riverso2013plug,betti2013distributed}).
Finally, note that coupling constraints among neighbours could in principle be incorporated in decentralized OCPs by finding a decentralized inner-approximation of the constraint sets. This, however, adds to the conservatism of decentralized OCPs.

In this paper, we make use of the OCP \eqref{sec2_dcn_OCP} to develop the proposed scheme which, at each time instant, performs two phases: a learning phase to learn the model uncertainty and an adaptation phase to compute the optimal control action. We assume that every subsystem can communicate with its neighbours at each time instant.  
While this assumption is not required for the adaptation phase which relies on the decentralized OCP \eqref{sec2_dcn_OCP} that requires no communication, it is still required for the learning phase. This is because the model uncertainty in \eqref{sec2_dcn_dyn2} relies on the states of the neighbours. Hence, communication is required among neighbours only once at each time instant in the learning phase. Since communication is allowed by assumption, the decentralized OCP \eqref{sec2_dcn_OCP} can in principle be replaced by a distributed OCP. While this may alleviate the conservatism of the developed scheme, the computation and communication complexity will increase significantly since distributed optimization is required.


In the learning phase, we aim to reduce the volume of the uncertainty set $\mathcal{W}_i$ in \eqref{sec2_dcn_dyn2} by learning the bounds $a_{ij}^{min}$ and $a_{ij}^{max}$ of the uncertain parameter $a_{ij}$ as more data is collected online. Hence, we develop a distributed set membership identification approach which learns these bounds by sharing information among the neighbours only once at each time instant. 
While the set $\mathcal{W}_i$ is a function of $a_{ij}^{max}$ only, $a_{ij}^{min}$ is also updated since it can speed up the shrinkage of  $\mathcal{W}_i$ by finding smaller upper bounds.
In the adaptation phase, we aim to adapt all MPC ingredients depending on the learnt uncertainty set and compute the local optimal control inputs taking the updated uncertainty set into consideration.
The adapted MPC ingredients include the rigid tube represented by $\mathcal{Z}_{K_i}$, the prestabilizing controller $K_i$, the tightened state constraint set $\mathcal{X}_i$ and the tightened input constraint set $\mathcal{U}_i$. 
Note that these ingredients are updated at each time instant $t$ based on the learnt uncertainty set $\mathcal{W}_i$ which is a function of the updated parameters $a_{ij}^{max}$ and $a_{ij}^{min}$ at this time instant. Hence, these MPC ingredients together with the uncertainty set are now functions of time.

In the sequel, we use the ellipsoidal rigid tube $\mathcal{Z}_{K_i}(t)=\{e_i : e_i^\top Z_i e_i \leq \alpha_i^2(t)\}$ and the linear prestabilizing controller $\kappa_i(x_i)=K_i(t)x_i$. We assume that $K_i(t)=K_{o_i}T_i(t)$ where $K_{o_i}$ is a full-rank stabilizing control gain and $T_i(t) \in \mathbb{S}^{n_i}_{++}$. Recall also that the tightened polytopic state and control constraint sets are given, respectively, by $\mathcal{X}_i(t) = \{x_i : G_i x_i \leq g_i(t)\}$ and $\mathcal{U}_i(t) = \{ u_i : H_i x_i \leq h_i(t)\}$. 
In addition to the decision variables in \eqref{sec2_dcn_OCP}, $\alpha_i(t)$, $T_i(t)$, $g_i(t)$ and $h_i(t)$ are also considered as decision variables in the developed scheme in order to update the MPC ingredients online; $K_{o_i}$, on the other hand, is assumed to be known in advance, computed, for example, with methods as those in \cite{aboudonia2021passivity}. 
When updating these ingredients at each time instant, one should ensure the following conditions,
\begin{enumerate}
    \vspace{-0.2cm}
    \item[I.] The set $\mathcal{Z}_{K_i}(t)$ is an RPI set for the dynamics \eqref{sec2_dcn_error}.
    \item[II.] The local matrices $A_{i}+B_iK_i(t)$ are Schur. 
    \item[III.] The global matrix $A+BK(t)$ is Schur.
    \item[IV.] The condition $\mathcal{X}_i(t) \bigoplus \mathcal{Z}_{K_i}(t) \subseteq \mathcal{Z}_i$ holds. 
    \item[V.] The condition $\mathcal{U}_i(t) \bigoplus K_i(t)\mathcal{Z}_{k_i}(t) \subseteq \mathcal{V}_i$ holds. 
    \vspace{-0.2cm}
\end{enumerate}
The aforementioned conditions are ensured by adding extra constraints in the OCP of the adaptation phase while preserving its convexity. These additional constraints are derived in the next section. 
Although the constraint sets are polytopic, the rigid-tube is chosen to be ellipsoidal. As shown in the next section, this leads in the adaptation phase to an OCP which considers the prestabilizing controller as a decision variable while maintaining the convexity of the OCP by making use of the S-lemma (\cite{boyd1994linear}). Recall also the scalability properties of ellipsoidal sets in previous related works when controlling networked dynamical systems (\cite{conte2016distributed,parsi2022scalable}).



The system \eqref{sec2_dcn_dyn} can be described by a graph $\mathcal{G}(\mathcal{M},\mathcal{E},\mathcal{A})$ where $\mathcal{M} = \{1, ..., M\}$ represents the set of nodes, $\mathcal{E} \subseteq (\mathcal{M} \times \mathcal{M})$ the set of edges and $\mathcal{A} = \{a_{ij} \in \mathbb{R} ,(i, j) \in \mathcal{E}\}$ the set of weights. Each subsystem is represented by a node. An edge exists from the $j$-th node to the $i$-th node if the $j$-th subsystem belongs to the set of in-neighbours of the $i$-th subsystem. We define the adjacency matrix $A_g \in \mathbb{R}^{M \times M}$ of the graph $\mathcal{G}$ such that the element in its $i$-th row and $j$-th column is given by $A_{g_{ij}}=a_{ij}$ if $i \in \mathcal{N}_j^+$ and zero otherwise. 
The degree matrix $D_g \in \mathbb{D}_{++}^{M}$ of the graph $\mathcal{G}$ is defined such that the diagonal element in the $i$-th row is given by $\sum_{j \in \mathcal{N}_i^-} a_{ij}$ and the Laplacian matrix is given by $L_g = D_g - A_g  \in \mathbb{S}_{+}^{M}$. 
These graph theory concepts are required to locally ensure Condition III that the global matrix $A+BK(t)$ is Schur. The global dynamics is then given by,
\begin{equation}
    \label{sec2_dyn_glb}
    z(t+1)=Az(t)+Bv(t)+d(t),
\end{equation}
where 
$z(t)=[z_1^\top(t),\hdots,z_M^\top(t)]^\top$, 
$v(t)=[v_1^\top(t),\hdots,$ $v_M^\top(t)]^\top$, 
$d(t)=[d_1^\top(t),\hdots,d_M^\top(t)]^\top$, 
$A_d=\operatorname{diag}(A_{1}, \hdots,$ $A_{M})$, 
$\tilde{A}_g = A_g \bigotimes I_{m_0}$,
$A=A_d+E \tilde{A}_g C$, 
$B=\operatorname{diag}(B_1,$ $\hdots,B_M)$, 
$C=\operatorname{diag}(C_1,\hdots,C_M)$ and 
$E=\operatorname{diag}(E_1,\hdots,$ $E_M)$.
Note that the proposed method is mainly developed for sparse graphs where only neighbours can communicate. While this method can be used with fully connected graphs where all subsystems can share information with each other, other efficient methods (e.g. \cite{lorenzen2019robust}) can be used where all subsystems share information with a central entity that does all the computations instead.
In the sequel, we make use of the following definitions.
\vspace{-0.25cm}
\begin{defn}[\cite{kouvaritakis2016model}]
    \label{def1}
    A set $Z_{K} \subset \mathbb{R}^{n}$ is robustly positively invariant under $z^+=Az+Bv$, $v=Kz$  and $z \in \mathcal{Z}$, $v \in \mathcal{V}$, if and only if $z^+ \in \mathcal{Z}_k$, $z \in \mathcal{Z}$, $v \in \mathcal{V}$ for all $z \in \mathcal{Z}_k$.
\end{defn}
\vspace{-0.25cm}
\begin{defn}[\cite{aliyu2017nonlinear}]
    \label{def2}
    The system $z^+=Az+Bv,$ $y=Cz+Dv$ is strictly passive if there exist functions $V_p:\mathcal{Z} \rightarrow \mathbb{R}$ and $\gamma_p:\mathcal{Z} \rightarrow \mathbb{R}_+$ so that $V_p(0)=0$ and $V_p(z^+)-V_p(z) \leq y^\top v - \gamma_p(z)$ for all $v \in \mathcal{V}$.
\end{defn}



\label{PF}

\section{Adaptive Learning-based MPC}

In this section, we start by discussing the distributed set membership identification approach used to learn the uncertainty set online in the learning phase. We then derive the constraints added to the local online OCP of each subsystem in the adaptation phase.
First, we discuss the learning phase where we aim to learn the uncertainty set $\mathcal{W}_i(t)$ in real time, as more data is collected online. According to \eqref{sec2_dcn_dyn2}, this set is a function of the bounds of the uncertain parameters. Hence, a better estimation of these bounds leads to a better description of this set. Thus, we develop here a distributed set-membership identification technique to improve the estimates of these bounds. Although such techniques are widely used with monolithic systems (e.g. see \cite{lorenzen2019robust}), we extend this technique here to interconnected systems with uncertain couplings where the identification is not carried out by a central entity.

Assume that the bounds of the uncertain parameter $a_{ij}$ at the $(t-1)$-th time instant are $a_{ij}^{min}(t-1)$ and $a_{ij}^{max}(t-1)$. Our goal is to improve the estimates of these bounds at the $t$-th time instant. For this purpose, we construct for every $a_{ij}$ the feasible parameter set defined by the inequality $F_{ij} a_{ij} \leq f_{ij}(t-1)$ where $F_{ij} = [1 \ -1]^\top$ and $f_{ij}(t-1) = [a_{ij}^{max}(t-1) \ a_{ij}^{min}(t-1)]^\top$. We then define the vector $\theta_i$ which comprises all the uncertain parameters to be identified by the $i$-th subsystem (i.e. $a_{ij}$, $j \in \mathcal{N}_i^-$). In this case, the feasible parameter set for the $i$-th subsystem is given by $F_i \theta_i \leq f_i(t-1)$
where $F_i$ is a block diagonal matrix with the matrices $F_{ij}$ for all $j \in \mathcal{N}_i^-$ on the diagonal and $f_i$ is a vector vertically concatenating the subvectors $f_{ij}$ for all $j \in \mathcal{N}_i^-$.

Recall that each subsystem can communicate with its neighbours in the learning phase. Hence, the states of the neighbours of the $i$-th subsystem at the $(t-1)$-th time instant are available for the $i$-th subsystem at the $t$-th time instant. Furthermore, the states and inputs of the $i$-th subsystem at the $(t-1)$-th time instant are available. Finally, the states of the $i$-th subsystem at the $t$-th time instant can be measured. Recall also that $\mathcal{D}_i = \{d_i : \Xi_i d_i \leq 1\}$. Defining $\chi_i(t-1)$ to be a matrix horizontally concatenating the vectors $C_j z_j(t-1)$ for all $j \in \mathcal{N}_i^-$, $\Omega_i(t-1)=-\Xi_i E_i \chi_i(t-1)$ and $\omega_i(t-1)=1 + \Xi_i A_{i} z_i(t-1) + \Xi_i B_i v_i(t-1) - \Xi_i z_i(t)$, we can construct for the $i$-th subsystem the non-falsified parameter set $\Delta_i = \{ \theta_i: \Omega_i(t-1) \theta_i \leq \omega_i(t-1) \}$.

In order to update the bounds of the uncertain parameters, the $i$-th subsystem should solve $2|\mathcal{N}_i^-|$ optimization problems at the $t$-th time instant where $|\cdot|$ denotes set cardinality. Each optimization problem updates one of these bounds. For each $j \in \mathcal{N}_i^-$, the lower and upper bounds of the parameter $a_{ij}$ are given respectively by,
\begin{subequations}
\vspace{-0.2cm}
\label{sec3_smid}
\begin{equation}
    \label{sec3_smid_min}
    a_{ij}^{min}(t) \hspace{-0.05cm}
    = \hspace{-0.05cm}
    \min_{a_{ij}} e_{ij}^\top \theta_i 
    \ \text{s.t.} \hspace{-0.1cm}
    \begin{bmatrix} 
        F_i \\ 
        \Omega_i(t-1) 
    \end{bmatrix} \hspace{-0.1cm}
    \theta_i \hspace{-0.1cm}
    \leq \hspace{-0.1cm}
    \begin{bmatrix} 
        f_i(t-1) \\ 
        \omega_i(t-1)
    \end{bmatrix}\hspace{-0.1cm},
\end{equation}
\vspace{-0.5cm}
\begin{equation}
    \label{sec3_smid_max}
    a_{ij}^{max}(t) \hspace{-0.1cm}
    = \hspace{-0.1cm}
    \max_{a_{ij}} 
    e_{ij}^\top \theta_i 
    \ \text{s.t.} \hspace{-0.1cm}
    \begin{bmatrix} 
        F_i \\ 
        \Omega_i(t-1) 
    \end{bmatrix} \hspace{-0.1cm}
    \theta_i \hspace{-0.1cm}
    \leq \hspace{-0.1cm} 
    \begin{bmatrix} 
        f_i(t-1) \\ 
        \omega_i(t-1)
    \end{bmatrix}\hspace{-0.1cm},
\end{equation}
\end{subequations}
where $e_{ij}$ is a unit vector which selects the parameter $a_{ij}$ in the vector $\theta_i$. 
Note that computing the new lower (upper) bound is a minimization (maximization) problem because we aim to find the smallest (largest) value of the corresponding parameter that lies inside the feasible parameter set and the non-falsified set.
In conclusion, the $i$-th subsystem learns the uncertainty set $\mathcal{W}_i(t)$ in \eqref{sec2_dcn_dyn2} by solving at the $t$-th time instant the optimization problems \eqref{sec3_smid} which identify the bounds of the uncertain parameter. These bounds are communicated among the in-neighbours and out-neighbours to be used in the adaptation phase. 
We also highlight the dependency of the upper bound $a_{ij}^{max}(t)$ on the lower bound $a_{ij}^{min}(t-1)$ which is part of the vector $f_i(t-1)$ in \eqref{sec3_smid_max} to support the fact that updating the lower bound might improve the updates of the upper bound.
Note that using asymmetric duality of linear programs, \eqref{sec3_smid} can be written compactly as a single optimization problem. We guarantee that the true parameters always lie within the updated bounds in the following theorem.
\begin{thm}
    Assume that $a_{ij}^{min}(0) \leq a_{ij} \leq a_{ij}^{max}(0)$. Then, $a_{ij}^{min}(t) \leq a_{ij} \leq a_{ij}^{max}(t)$ for all $t \geq 0$ where $a_{ij}^{min}(t)$ and $a_{ij}^{max}(t)$ are computed using \eqref{sec3_smid}.
    \vspace{-0.4cm}
\end{thm}
\begin{pf}
    Note that $z_i(1) = A_{i}z_i(0)+B_iv_i(0) + E_i \sum_{j \in \mathcal{N}_i^-} a_{ij} C_j z_j(0) + d_i(0)$ and recall that $d_i(0) \in \mathcal{D}_i$ for all $i \in \mathcal{M}$. 
    Thus, the true parameters satisfy $\Xi z_i(1) - \Xi A_{i} z_i(0) + \Xi B_i v_i(0) + \Xi E_i \sum_{j \in \mathcal{N}_i^-} a_{ij} C_j z_j(0)) \leq 1$ and hence, $\Omega_i(0) \theta_i \leq \omega(0)$ for all $i \in \mathcal{M}$.
    Since $a_{ij}^{min}(0) \leq a_{ij} \leq a_{ij}^{max}(0)$, the true parameters also satisfy $F_i \theta_i \leq f_i(0)$ for all $i \in \mathcal{M}$ by assumption.
    Since $a_{ij}^{min}(1)$ and $a_{ij}^{max}(1)$ are computed using \eqref{sec3_smid}, then these are the smallest and largest values respectively satisfying $\Omega_i(0) \theta_i \leq \omega(0)$ and $F_i \theta_i \leq f_i(0)$ that the true parameters also satisfy and thus, $a_{ij}^{min}(1) \leq a_{ij} \leq a_{ij}^{max}(1)$. 
    By induction, the same results hold for all $t$.
    \hfill
    $\square$
    \vspace{-0.4cm}
\end{pf}
As a consequence of the above theorem, it is guaranteed that $a_{ij}^{min}(t) \geq a_{ij}^{min}(t-1)$ because $a_{ij}^{min}(t)$ is computed by minimizing over all possible values satisfying $a_{ij} \geq a_{ij}^{min}(t-1)$. The same arguments apply to the upper bound and hence, $\mathcal{W}_i(t) \subseteq \mathcal{W}_i(t-1)$ for all $t$ (i.e. the volume of the uncertainty set can never increase).
Note that the proposed scheme is not necessarily restricted to set membership identification. While this technique is used here to update uncertain parameters in linear coupling terms, the proposed scheme paves the way to develop other schemes which can learn more complex uncertainties with nonlinear couplings and other disturbances. In this case, one can use other machine learning methods which have shown success in learning uncertainty sets.
We now move to discussing the adaptation phase.
As mentioned in Section \ref{PF}, the online OCP is solved in the adaptation phase to adapt the MPC ingredients based on the learnt uncertainty set and compute the optimal control inputs. To adapt the MPC ingredients online while ensuring robust constraint satisfaction, the five conditions mentioned in Section \ref{PF} should be satisfied. We derive here additional constraints which ensure the satisfaction of these conditions while maintaining the convexity of the OCP \eqref{sec2_dcn_OCP}.

We start with the condition which ensures the robust positive invariance of the set $\mathcal{Z}_{K_i}(t)$. First, we define for the $i$-th subsystem the output vector $y_i(t) = C_i z_i(t)$ and the constraint set $\mathcal{Y}_i$, obtained from $\mathcal{Z}_i$ in the obvious way, such that $y_i \in \mathcal{Y}_i$. We then compute a minimum volume ellipsoidal set $\hat{\mathcal{Y}}_i = \{y_i : \|y_i-c_{y_i}\|^2_{Y_i} \leq 1\}$ that contains the set $\mathcal{Y}_i$. For this purpose, the vertex description of $\mathcal{Y}_i$ is needed. Note that this is in general combinatorial in the linear inequalities needed to define $\mathcal{Y}_i$. However, the sets $\mathcal{Y}_i$ of all subsystems do not change over time and hence, their minimum volume ellipsoidal sets $\hat{\mathcal{Y}}_i$ can be computed once offline by solving a convex optimization problem for each. Finally, we construct the outer approximation $\hat{\mathcal{W}}_i(t) = \bigoplus_{j \in \mathcal{N}_i^-} a_{ij}^{max}(t) \hat{\mathcal{Y}_j}$ of the uncertainty set $\mathcal{W}_i(t) = \bigoplus_{j \in \mathcal{N}_i^-} a_{ij}^{max}(t) \mathcal{Y}_j$. Note that  $\hat{\mathcal{W}}_i(t)$ is not necessarily an ellipsoidal set, but it is the Minkowski sum of the ellipsoidal sets $\mathcal{W}_{ij}(t) = a_{ij}^{max}(t)\hat{\mathcal{Y}}_j = \{w_{ij} : \|w_{ij}-a_{ij}^{max}(t)c_{y_j}\|_{Y_j}^2 \leq a_{ij}^{max^2}(t)\}$. 
Similarly, we also compute a minimum volume ellipsoidal set $\hat{\mathcal{D}}_i = \{d_i : \|d_i\|_{{\hat{\Xi}}_i}^2 \leq 1\}$ offline for the set $\mathcal{D}_i$ of the $i$-th subsystem for all $i \in \{1,\hdots,M\}$. 

In Proposition \ref{sec3_prop_RPI}, we consider Condition I in Section \ref{PF} and derive constraints which ensure the robust positive invariance of the set $\mathcal{Z}_{K_i}(t)$ for the error dynamics \eqref{sec2_dcn_error} under the disturbances $w_i \in \hat{\mathcal{W}}_i(t)$ and $d_i \in \hat{\mathcal{D}}_i$. Although this might be more conservative than considering the disturbances $w_i \in \mathcal{W}_i(t)$ and $d_i \in \mathcal{D}_i$, the convexity of the resulting OCP is preserved in this case. Another source of conservatism in Proposition \ref{sec3_prop_RPI} is using the approximate S-lemma to reach local convex constraints. Recall here that $K_i(t) = K_{o_i} T_i(t)$. The dependence on time in \eqref{sec3_mrpi1} is omitted in the interest in space

\begin{prop}
    \label{sec3_prop_RPI}
    The set $Z_{K_i}(t)$ is robustly positively invariant for the error dynamics \eqref{sec2_dcn_error} with the uncertainty sets $\hat{\mathcal{W}}_i(t)$ and $\hat{\mathcal{D}_i}$ if there exist non-negative scalars $\lambda_{i}(t)$, $\lambda_{d_i}(t)$  and $\lambda_{ij}(t)$ for all $j \in \mathcal{N}_i^-$ such that \eqref{sec3_mrpi1} holds.
    \vspace{-0.4cm}
\end{prop}

\begin{table*}
	\normalsize
	\begin{equation}
	\label{sec3_mrpi1}
	\left[
    \begin{array}{c;{2pt/2pt}c;{2pt/2pt}c;{2pt/2pt}c;{2pt/2pt}c}
	    \lambda_{i} Z_i & 0 & 0 & 0 & A_{i} \alpha_i + B_i K_{o_i} T_i \alpha_i^\top \\ \hdashline[2pt/2pt]
	    * & Y_{\lambda_i} & 0 & -Y_{\lambda_i}\tau_i & \mathcal{E}_i^\top \\ \hdashline[2pt/2pt]
	    * & * & \lambda_{d_i} \hat{\Xi}_i & 0 & I_{n_i} \\ \hdashline[2pt/2pt]
	    * & * & * & \alpha_i - \lambda_i - \lambda_{d_i} - \sum\nolimits_{j \in \mathcal{N}_i^-} \lambda_{ij} (1-\|c_{y_j}\|_{Y_j}^2) a_{ij}^{max^2} & 0 \\ \hdashline[2pt/2pt]
	    * & * & * & * & \alpha_i Z_i^{-1} \\
	\end{array}
	\right]
	\geq 0.
    \vspace{-0.2cm}
	\end{equation}
\end{table*}

\begin{pf}
    According to Definition \ref{def1}, the set $Z_{K_i}(t)$ is robustly positively invariant under the considered dynamics and uncertainty sets if and only if $\|e_i^+\|_{Z_i}^2 \leq \alpha_i^2(t)$ for all $e_i$ satisfying $\|e_i\|_{Z_i}^2 \leq \alpha_i^2(t)$, $w_{ij}$ satisfying $\|w_{ij}-a_{ij}^{max}(t) c_{y_j}\|_{Y_j}^2 \leq a_{ij}^{{max}^2}(t)$ for all $j \in \mathcal{N}_i^-$ and $d_i$ satisfying $\|d_i\|_{\hat{\Xi}_i}^2 \leq 1$. 
    By defining $s_i$ such that $e_i = \alpha_i(t) s_i$, the invariance condition becomes $\|(A_{i} \alpha_i(t) + B_i K_{o_i} T_i(t) \alpha_i(t)) s_i + E_i w_i + d_i\|_{Z_i}^2 \leq \alpha_i^2(t)$ for all $s_i$ satisfying $\|s_i\|_{Z_i}^2 \leq 1$, $w_{ij}$ satisfying $\|w_{ij}-a_{ij}^{max}(t) c_{y_j}\|_{Y_j}^2 \leq a_{ij}^{{max}^2}(t)$ for all $j \in \mathcal{N}_i^-$ and $d_i$ satisfying $\|d_i\|_{\hat{\Xi}_i}^2 \leq 1$. 
    We now define the block diagonal matrix $Y_{\lambda_i(t)}$ with the submatrices $\lambda_{ij}(t) Y_j$ for all $j \in \mathcal{N}_i^-$ on the diagonal and the vector $\tau_i(t)$ which vertically concatenates the subvectors $a_{ij}^{max}(t) c_{y_j}$. 
    Using the S-lemma and applying the Schur complement yields \eqref{sec3_mrpi1} where $I_{n_i}$ is an identity matrix of size $n_i$ and $\mathcal{E}_i$ is a matrix horizontally concatenating the matrix $E_i$ for $|\mathcal{N}_i^-|$ times.
    \hfill
    $\square$
\end{pf}



Note that the dimension of the constraint in \eqref{sec3_mrpi1} increases linearly with respect to the number of neighbours. 
Rigid tubes are known for their little computation burden in comparison to homothetic tubes (\cite{rakovic2012homothetic}). Since we use time-varying rigid tubes, one might think that rigid tubes lose their main computational advantage and could be better replaced by homothetic tubes. This is, however, not the case because when using homothetic tubes, the nominal dynamics would be replaced by tube dynamics. Since the prestabilizing controller is a decision variable and the tube is ellipsoidal, the tube dynamics is represented by one matrix inequality for each prediction step. In this case, the problem becomes computationally challenging as the prediction horizon gets longer.

Next, we consider Conditions II and III in Section 2 by deriving constraints ensuring that $A_{i}+B_iK_i(t)$ for all $i \in \{1,\hdots,M\}$ and $A+BK(t)$ are Schur. 
For this purpose, we rely on passivity arguments for discrete-time interconnected systems. 
First, we derive a constraint ensuring that the noise-free dynamics of the $i$-th subsystem (i.e. \eqref{sec2_dcn_dyn} with $d_i(t)=0$ for all $t$) under the controller $u_i(t)=K_{o_i}T_i(t)x_i(t)$ is strictly passive. We show that this implies that $A_{i}+B_iK_i(t)$ is Schur. We then derive other constraints ensuring that the strict passivity of all subsystems implies the asymptotic stability of the whole system ensuring that $A+BK(t)$ is Schur. 
To ensure strict passivity of the $i$-th subsystem, we define the virtual output $\tilde{y}_i(t)=C_iz_i(t)+D_iw_i(t)$ where $D_i \in \mathbb{D}_{++}^{m_0}$ is a decision variable. We also assume that a matrix $P_i \in \mathbb{S}_{++}^{n_i}$ is computed offline together with $K_{o_i}$, for example as in \cite{aboudonia2021passivity}, so that $V_{p_i}(z_i)=\|z_i\|_{P_i}^2$ is a storage function over the set $\mathcal{Z}_i$. We finally define a minimum volume ellipsoidal set $\hat{\mathcal{Z}}_i=\{z_i:\|z_i-c_{z_i}\|_{\hat{Z}_i} \leq 1\}$ for the state constraint set $\mathcal{Z}_i$. The dependence on time in \eqref{sec3_psv1} is omitted in the interest in space.

\begin{prop}
    \label{sec3_prop2}
    The noise-free dynamics of the $i$-th subsystem $z_i^+=(A_{i}+B_iK_{o_i}T_i(t))z_i + E_iw_i$ is strictly passive with respect to $w_i \in \hat{\mathcal{W}}_i(t)$ and $\tilde{y}_i \in C_i\hat{\mathcal{Z}}_i \bigoplus D_i \hat{\mathcal{W}}_i(t)$ if there exist $\Gamma_i(t)  \in \mathbb{D}_{++}^{n_i}$, $D_i(t)  \in \mathbb{D}_{++}^{n_i}$, $T_i(t) \in \mathbb{S}_{++}^{n_i}$, $\alpha_i(t) > 0$ such that,
    \begin{equation}
	    \label{sec3_psv1}
        \left[
        \hspace{-0.05cm}
        \begin{array}{c;{2pt/2pt}c;{2pt/2pt}c}
            (P_i - \Gamma_i) \alpha_i & 0.5 C_i^\top \alpha_i & (A_{i}\alpha_i+B_i K_{o_i}T_i\alpha_i)^\top \\ \hdashline[2pt/2pt]
            * & D_i \alpha_i & E_i^\top \alpha_i \\ \hdashline[2pt/2pt]
            * & * &
            P_i^{-1} \alpha_i 
        \end{array}
        \hspace{-0.06cm}
        \right]
        \hspace{-0.15cm} \geq \hspace{-0.05cm} 0.
    \end{equation}
\end{prop}
\begin{pf} 
    To ensure strict passivity, we have to satisfy the condition in Definition \ref{def2} for all $w_i \in \hat{\mathcal{W}}_i(t)$. Recall also that the storage function $V_{p_i}(z_i)$ is defined over the set $\hat{\mathcal{Z}}_i$.
    Hence, the noise-free dynamics of the $i$-th subsystem under the controller $u_i(t)=K_{o_i}T_i(t)x_i(t)$ is strictly passive according to Definition \ref{def2} if $\|z_i^+\|_{P_i}^2 - \|z_i\|_{P_i}^2 \leq \tilde{y}_i^\top w_i - \|z_i\|_{\Gamma_i}^2$ for all $z_i$ satisfying $\|z_i-c_{z_i}\|_{\hat{Z}_i}^2 \leq 1$ and $w_{ij}$ satisfying $\|w_{ij}-a_{ij}^{max}(t) c_{y_j}\|_{Y_j}^2 \leq a_{ij}^{{max}^2}(t)$ for all $j \in \mathcal{N}_i^-$. 
    By substituting for the dynamics, the strict passivity condition is given by $\|z_i\|_{L_{1_i}}^2 + 2 w_i^\top L_{2_i} z_i + \|w_i\|_{L_{3_i}}^2 \geq 0$ for all $z_i$ satisfying $\|z_i-c_{z_i}\|_{\hat{Z}_i}^2 \leq 1$ and $w_{ij}$ satisfying $\|w_{ij}-a_{ij}^{max}(t) c_{y_j}\|_{Y_j}^2 \leq a_{ij}^{{max}^2}(t)$ for all $j \in \mathcal{N}_i^-$ where $L_{1_i}=P_i-\Gamma_i-(A_{i}+B_i K_{o_i} T_i(t))^\top P_i (A_{i}+B_i K_{o_i} T_i(t))$, $L_{2_i}=0.5C_i-E_i^\top P_i (A_{i}+B_i K_{o_i} T_i(t))$ and $L_{3_i}=D_i-E_i^\top P_i E_i$.
    We now define the block diagonal matrix $Y_{\sigma_i(t)}$ with the submatrices $\sigma_{ij}(t) Y_j$ for all $j \in \mathcal{N}_i^-$ on the diagonal and the vector $\tau_i(t)$ which vertically concatenates the subvectors $a_{ij}^{max}(t) c_{y_j}$ for all $j \in \mathcal{N}_i^-$. 
    We also recall that the ellipsoidal set $\hat{\mathcal{Y}}_i$ contains the origin in its interior and hence, $\|c_{y_i}\|_{Y_i}^2 \leq 1$.
    Using the S-lemma, applying the Schur complement and multiplying the resulting matrix inequality by $\alpha_i(t)$ thus leads to \eqref{sec3_psv1}.
    \vspace{-0.2cm}
    \hfill
    $\square$
\end{pf}
Note that \eqref{sec3_psv1} implies that $ P_i-(A_{i}+B_i K_{o_i} T_i(t))^\top P_i (A_{i} + B_i K_{o_i} T_i(t)) \geq \Gamma_i(t)$ where $\Gamma_i(t) \in \mathbb{D}_{++}^{n_i}$ and hence, $A_{i}+B_iK_i(t)$ is Schur. 
One source of conservatism in Proposition \ref{sec3_prop2} is that the local controller is designed to not only stabilize the local dynamics, but also to passivize it. The benefit of passivizing the local dynamics will become obvious in the next proposition.
We now consider Condition III in Section 2 by deriving constraints which ensure that the strict passivity of all noise-free subsystems implies the asymptotic stability of the overall noise-free system. For this purpose, we consider the noise-free global dynamics
$z(t+1) = A_{cl}(t) z(t) + E w(t)$
where $A_{cl}(t)=\operatorname{diag}(A_{1}+B_1K_{1}(t),\hdots,A_{M}+B_MK_{M}(t))$, $E=\operatorname{diag}(E_1,\hdots,E_M)$ and $w(t)=[w_1^\top(t),\hdots,w_M^\top(t)]^\top$.
We also define the Lyapunov function $V(z) = \|z\|_P^2$ where $P=\operatorname{diag}(P_1,...,P_M)$. Furthermore, we define the matrices $\Gamma(t) = \operatorname{diag}(\Gamma_1(t),...,\Gamma_M(t))$, $C = \operatorname{diag}(C_1,...,C_M)$ and $D(t) = \operatorname{diag}(D_1(t),...,D_M(t))$. 
Recall that $A_g$ and $D_g$ are the adjacency and degree matrices, respectively. The global virtual output vector is given by $\tilde{y}(t) = C z(t) + D w(t)$ and the global coupling vector is given by $w(t) = \tilde{A}_g C z(t)$. 

Finally, we construct $A_g^{max}(t)$ and $D_g^{max}(t)$ by replacing all uncertain parameters in $A_g$ and $D_g$ with their upper bounds. Consequently, we define $U=\operatorname{diag}(U_1,...,U_M)=C^\top D_g C$, $V=[V_1^\top,...,V_M^\top]^\top=C^\top \tilde{A}_g^\top$ and $W=[W_1^\top,...,W_M^\top]^\top=\tilde{A}_g C$ where $U_i \in \mathbb{R}^{n_i \times n_i}$, $V_i \in \mathbb{R}^{n_i \times m_0}$ and $W_i \in \mathbb{R}^{m_0 \times n_i}$. We also construct $U^{max}(t)$, $V^{max}(t)$ and $W^{max}(t)$ by replacing all uncertain parameters in $U$, $V$ and $W$ with their upper bounds. In the sequel, we denote the diagonal element in the $j$-th row of a matrix by $[\cdot]_j$ and the summation of the absolute values of the entries in the $j$-th row of a matrix by $|\cdot|_j$. The dependence on time in \eqref{sec3_st4} is omitted in the interest in space.

\begin{prop}
    \label{prop3}
    The strict passivity of the noise-free dynamics of all subsystems under the corresponding controller $u_i(t)=K_{o_i}T_i(t)x_i(t)$ $(\text{i.e. } z_i(t+1)=(A_{i}+B_iK_{o_i}T_i(t))z_i(t) + E_iw_i(t))$ implies the asymptotic stability of the noise-free dynamics of the overall system $(\text{i.e. } z(t+1) = A_{cl}(t) z(t) + E w(t))$ if,
    \begin{equation*}
        (\left[\Gamma_i\right]_j-\epsilon) \alpha_i \geq |U_i^{max}|_j \alpha_i + |V_i^{max}|_j \alpha_i, \forall \ j \in \{1,...,n_i\},
    \end{equation*}
    \vspace{-0.6cm}
    \begin{equation}
        \label{sec3_st4}
        [D_i]_j \alpha_i \leq \alpha_i/|W_i^{max}|_j, \ \forall j \in \{1,...,m_0\} :  |W_i^{max}|_j \neq 0. %
    \end{equation}
    \vspace{-0.75cm}
\end{prop}

\begin{pf}
    The passivity of the noise-free dynamics of each subsystem implies that $\| z_i^+\|_{P_i}^2-\| z_i\|_{P_i}^2 \leq \tilde{y}_i^\top w_i - \|z_i\|_{\Gamma_i(t)}^2$. Summing up these inequalities for all subsystems yields $V(z^+)-V(z) \leq \tilde{y}^\top w - \|z\|_{\Gamma(t)}$. To ensure asymptotic stability, it suffices to guarantee that $\tilde{y}^\top w - \|z\|_{\Gamma(t)} < 0$. Using the expressions of $\tilde{y}$ and $w$, this condition can be written as $\|z\|^2_{\Gamma(t) - C^\top \tilde{A}_g C - C^\top \tilde{A}_g^\top D(t) \tilde{A}_g C} \geq \epsilon \|z\|^2$ where $\epsilon > 0$. Recall that $L_g = D_g - A_g$ and hence, $\|z\|^2_{\Gamma(t) + C^\top L_g C - C^\top D_g C - C^\top \tilde{A}_g^\top D(t) \tilde{A}_g C} \geq \epsilon \|z\|^2$. Note also that $L_g$ is positive semidefinite by definition and hence, it is sufficient to ensure that $\|z\|^2_{\Gamma(t) - C^\top D_g C - C^\top \tilde{A}_g^\top D(t) \tilde{A}_g C} \geq \epsilon \|z\|^2$. This condition can be reformulated using the Schur Complement as,
    \begin{equation}
        \label{sec3_st3}
        \left[
        \begin{array}{c;{2pt/2pt}c}
            \Gamma(t) - C^\top D_g C -\epsilon I_{n}& C^\top \tilde{A}_g^\top \\ \hdashline[2pt/2pt]
            * & D(T)^{-1} \\
        \end{array}
        \right]
        \geq 0.
    \end{equation}
    Using diagonal dominance, the constraint \eqref{sec3_st3} is approximated for each subsystem by $[\Gamma_i(t)]_j -\epsilon \geq |U_i|_j + |V_i|_j$ for all $j \in \{1,...,n_i\}$ and $[D_i(t)^{-1}]_j \geq |W_i|_j$ for all $j \in \{1,...,m_0\}$. Multiplying the former by $\alpha_i(t)$ and the latter by $\alpha_i(t)^{-1}$ and inverting, these inequalities become $(\left[\Gamma_i(t)\right]_j-\epsilon) \alpha_i(t) \geq |U_i|_j \alpha_i(t) + |V_i|_j \alpha_i(t)$ for all $j \in \{1,...,n_i\}$ and
    $[D_i(t)]_j \alpha_i(t) \leq \alpha_i(t)/|W_i|_j$ for all $j \in \{1,...,m_0\}$ such that $|W_i|_j \neq 0$. Note that the resulting inequalities are functions of the uncertain parameters. More conservative inequalities can be obtained by replacing the uncertain parameters with their upper bounds, leading to \eqref{sec3_st4}. 
    \hfill
    $\square$
\end{pf}

In Proposition \ref{prop3}, \eqref{sec3_st4} ensures that the Lyapunov condition is satisfied for the noise-free global dynamics $z(t+1)=A_{cl}(t)z(t)+Ew(t)$ or equivalently, $z(t+1)=(A+BK(t))z(t)$ and hence, $A+BK(t)$ is Schur. Sources of conservatism in this proposition include implying $\tilde{y}^\top w - \|z\|_{\Gamma(t)} < 0$ instead of $V(z^+)-V(z) < 0$, ignoring the term $C^\top L_g C$, using diagonal dominance and replacing the true values of the parameters with their upper bounds. These sources are utilized to eventually reach local convex constraints.
We now move to Conditions IV and V in Section 2.
In the sequel, we denote the $j$-th row of the matrices $G_i$ and $H_i$ by $G_i^{(j)}$ and $H_i^{(j)}$ respectively.
The dependence on time is omitted in \eqref{sec3_mku0} and \eqref{sec3_mku3} in the interest in space.

\begin{prop}
    \label{prop4}
    The Minkowski sum of $\mathcal{X}_i(t)$ and $\mathcal{Z}_{K_i}(t)$ is contained in $\mathcal{Z}_i$ if there exist $\mu_{ij}(t) \geq 0$ for all $j \in \{1,...,q_i\}$ such that,
    \begin{equation}
    \label{sec3_mkx}
        \left[
        \begin{array}{c;{2pt/2pt}c}
            \mu_{ij}(t)Z_i & -0.5G_i^{(j)^\top} \alpha_i(t) \\ \hdashline[2pt/2pt]
            * & 1-\mu_{i_j}(t)-g_i(t)
        \end{array}
        \right]
        \geq 0.
    \end{equation}
\end{prop}

\begin{pf}
    The Minkowski sum of $\mathcal{X}_i(t)$ and $\mathcal{Z}_{K_i}(t)$ is contained in $\mathcal{Z}_i$ if and only if for all $j \in \{1,...,q_i\}$, $G_i^{(j)} (x_{e_i}+x_{s_i}) \leq 1$ for all $x_{e_i}$ such that $\|x_{e_i}\|_{Z_i}^2 \leq \alpha_i^2(t)$ and for all $x_{s_i}$ such that $G_i x_{s_i} \leq g_i(t)$. 
    By defining $s_i$ and $t_i$ such that $x_{e_i}=\alpha_i(t) s_i$ and $x_{s_i}=g_i(t) t_i$, this condition becomes $G_i^{(j)} \alpha_i(t) s_i + G_i^{(j)} g_i(t) t_i \leq 1$ for all $j \in \{1,...,q_i\}$, $s_i$ satisfying $s_i^\top Z_i s_i \leq 1$ and $t_i$ satisfying $G_i t_i \leq 1$. Using the S-lemma, this condition is given for all $j \in \{1,...,q_i\}$ by,
    \begin{equation}
        \label{sec3_mkx1}
        \left[
        \begin{array}{c;{2pt/2pt}c;{2pt/2pt}c}
            \mu_{ij}(t) Z_i & 0 & - 0.5 \alpha_i(t) G_i^{(j)^\top} \\ \hdashline[2pt/2pt]
            * & 0 & \sum_{l=1}^{q_i} 0.5 \mu_{ijl}(t) G_i^{(l)^\top} - 0.5 G_i^{(j)^\top} g_i(t)\\ \hdashline[2pt/2pt]
            * & * & 1-\mu_{ij}(t)-\sum_{l=1}^{q_i} \mu_{ijl}(t) \\
        \end{array}
        \right]
        \hspace{-0.15cm} \geq \hspace{-0.05cm} 0.
    \end{equation}    
    Note that \eqref{sec3_mkx1} holds if $\sum_{l=1}^{q_i} 0.5 \mu_{ijl}(t) G_i^{(l)} = 0.5 G_i^{(j)} g_i(t)$. By stacking this equality for all $j \in \{1,...,q_i\}$, we reach $g_i(t) G_i = \mu_i(t) G_i$ where $\mu_{ijl}(t)$ is the entry of $\mu_i(t)$ in the $j$-th row and $l$-th column. To satisfy this equality, we choose the diagonal elements of $\mu_i(t)$ to be equal to $g_i(t)$ and the off-diagonal elements to be equal to zero and hence, \eqref{sec3_mkx1} reduces to \eqref{sec3_mkx}.
    \hfill
    $\square$
\end{pf}

\begin{prop}
    \label{prop5}
    The Minkowski sum of $\mathcal{U}_i(t)$ and $K_{o_i}T_i(t)\mathcal{Z}_{k_i}(t)$ is contained in $\mathcal{V}_i$ if there exist $\nu_{ij}(t) > 0$ for all $j \in \{1,...,r_i\}$ and $M_i(t) \in \mathbb{S}^{p_i}_{++}$ such that,
    \begin{equation}
        \label{sec3_mku0}
        \left[
        \begin{array}{c;{2pt/2pt}c}
            M_i^{-1} & K_i^\top \alpha_i \\ \hdashline[2pt/2pt]
            * & \nu_{ij} Z_i \\
        \end{array}
        \right]
        \hspace{-0.05cm} \geq \hspace{-0.05cm} 0,
        \left[
        \begin{array}{c;{2pt/2pt}c}
            M_i^{-1} & 0.5 M_i^{-1} H_i^{(j)^\top} \\ \hdashline[2pt/2pt]
            * & 1 - \nu_{ij} - h_i \\
        \end{array}
        \right]
        \hspace{-0.05cm} \geq \hspace{-0.05cm} 0.
    \end{equation}
\end{prop}

\begin{pf}
    First, we define $M_i(t) \in \mathbb{S}_{++}^{p_i}$ and $\nu_{ij}(t) > 0$ such that $M_i(t) \leq \nu_{ij}(K_{o_i} T_i(t) \alpha_i(t) Z_i^{-1} \alpha_i(t) T_i^\top(t) K_{o_i}^\top)^{-1}$ for all $j \in \{1,...,r_i\}$. Since $T_i(t) \in \mathbb{S}_{++}^{n_i}$, $\alpha_i(t) > 0$ and $K_{o_i}$ is full-rank, this inequality is invertible and hence, $M_i(t)^{-1} - \nu_{ij}(t)^{-1} K_{o_i} T_i(t) \alpha_i(t) Z_i^{-1} \alpha_i(t) T_i^\top(t) K_{o_i}^\top \alpha_i(t) \geq 0$  for all $j \in \{1,...,r_i\}$.
    Applying the Schur Complement yields the first condition in \eqref{sec3_mku0}. 
    The Minkowski sum of $\mathcal{U}_i(t)$ and $K_{o_i}T_i(t)\mathcal{Z}_{K_i}(t)$ is contained in $\mathcal{V}_i$ if and only if for all $j \in \{1,...,r_i\}$, $H_i^{(j)} (u_{e_i}+u_{s_i}) \leq 1$ for all $u_{e_i}$ satisfying $\|u_{e_i}\|_{({K_{o_i} T_i(t) Z_i^{-1} T_i^\top(t) K_{o_i}^\top})^{-1}}^2 \leq \alpha_i^2(t)$ and $u_{s_i}$ satisfying $H_i u_{s_i} \leq h_i(t)$. 
    By defining $s_i$ and $t_i$ such that $u_{e_i}= s_i$ and $u_{s_i}=h_i(t) t_i$, we reach the condition $H_i^{(j)} s_i + H_i^{(j)} h_i(t) t_i \leq 1$ for all $j \in \{1,...,r_i\}$, $s_i$ and $t_i$ satisfying $s_i^\top (K_{o_i} T_i(t) \alpha_i(t) Z_i^{-1} \alpha_i(t) T_i^\top(t) K_{o_i}^\top)^{-1} s_i \leq 1$ and $H_i t_i \leq 1$. 
    Using the S-lemma, we reach  for all $j \in \{1,...,r_i\}$, 
    \begin{equation}
        \label{sec3_mku3}
        \left[
        \begin{array}{c;{2pt/2pt}c;{2pt/2pt}c}
            \nu_{ij} \psi_i^{-1} & 0 & - 0.5 H_i^{(j)^\top} \\ \hdashline[2pt/2pt]
            * & 0 & \sum_{l=1}^{r_i} 0.5 \nu_{ijl} H_i^{(l)^\top} - 0.5 H_i^{(j)^\top} h_i(t)\\ \hdashline[2pt/2pt]
            * & * & 1-\nu_{ij}-\sum_{l=1}^{r_i} \nu_{ijl} \\
        \end{array}
        \right]
        \hspace{-0.15cm} \geq \hspace{-0.05cm} 0,
    \end{equation}
    where $\psi_i(t) = K_{o_i} T_i(t) \alpha_i(t) Z_i^{-1} T_i^\top K_{o_i}^\top(t) \alpha_i(t)$.
    Note that \eqref{sec3_mku3} holds if $\sum_{l=1}^{r_i} 0.5 \nu_{ijl}(t) H_i^{(l)} = 0.5 H_i^{(j)^\top} h_i(t)$. 
    By stacking this equality for all $j \in \{1,...,r_i\}$, we reach $h_i(t) H_i = \nu_i(t) H_i$ where $\nu_{ijl}(t)$ is the entry of $\nu_i(t)$ in the $j$-th row and $l$-th column. 
    To satisfy this equality, we choose the diagonal elements of $\nu_i(t)$ to equal $h_i(t)$ and the off-diagonal elements to equal zero. 
    Recalling that $M_i(t) \leq \nu_{ij}(t) (K_i(t) \alpha_i(t) Z_i^{-1} K_i^\top(t) \alpha_i(t))^{-1}$, \eqref{sec3_mku3} reduces to a matrix, that if pre and post-multiplied by $\operatorname{diag}(M_i(t)^{-1},1)$, yields the second condition in \eqref{sec3_mku0}.
    \hfill
    $\square$
\end{pf}


One source of conservatism in Proposition \ref{prop4} is choosing the diagonal elements of $\mu_i(t)$ to equal $g_i(t)$ and the off-digonal elements to equal zero. The same applies to $\nu_i(t)$ in Proposition \ref{prop5}. Additionally, introducing $M_i(t)$ adds one more degree of conservatism in Proposition \ref{prop5}.
In conclusion, the OCP of the $i$-th subsystem becomes,
\begin{equation}
    \label{OCP}
    \begin{aligned}
        \min 
        J_i 
        \text{ s.t.} \left\{
        \begin{aligned}
            & \forall k \in \{0,...,N-1\}, \\
            & z_i(t) - x_i(0|t) \in \mathcal{Z}_{K_i} (t), \\
            & x_i(k+1|t)=A_{i}x_i(k|t)+B_iu_i(k|t), \\
            & x_i(k|t) \in \mathcal{X}_{i}(t), \quad u_i(k|t) \in \mathcal{U}_{i}(t), \\
            & x_i(N|t) = x_{e_i}(t)=A_{i}x_{e_i}(t)+B_iu_{e_i}(t), \\
            & x_{e_i} \in \xi \mathcal{X}_{i}(t), \quad u_{e_i} \in \xi \mathcal{U}_{i}(t), \\
            & \eqref{sec3_mrpi1}, \ \lambda_{i}(t), \ \lambda_{d_i}(t), \ \lambda_{ij}(t) \geq 0, \ \forall j \in \mathcal{N}_i^-,\\
            & \eqref{sec3_psv1}, \eqref{sec3_st4},  \eqref{sec3_mkx}, \ \mu_{ij}(t) \geq 0 \ \forall j \in \{1,...,q_i\}, \\
            & \eqref{sec3_mku0}, \ \nu_{ij}(t) > 0 \ \forall j \in \{1,...,r_i\}, \\
            & M_i(t) \in \mathbb{S}_{++}^{n_i}, \ T_i(t) \alpha_i(t) \in \mathbb{S}_{++}^{n_i}, \\
            & \Gamma_i(t) \alpha_i(t) \in \mathbb{D}_{++}^{n_i}, \ D_i(t) \alpha_i(t) \in \mathbb{D}_{++}^{n_i}.
        \end{aligned}
        \right.
    \end{aligned}
\end{equation}
The decision variables are $x_i(k|t)$ and $u_i(k|t)$ for all $k \in \{0,...,N\}$, $x_{e_i}(t)$, $u_{e_i}(t)$, $g_i(t)$, $h_i(t)$, $\alpha_i(t)$, $T_i(t) \alpha_i(t)$, $\Gamma_i(t) \alpha_i(t)$, $D_i(t) \alpha_i(t)$, $M_i(t)^{-1}$, $\lambda_{i}(t)$, $\lambda_{d_i}(t)$, $\lambda_{ij}(t)$ for all $j \in \mathcal{N}_i^-$, $\mu(t)_{ij}$ for all $j \in \{1,...,q_i\}$ and $\nu(t)_{ij}$ for all $j \in \{1,...,r_i\}$. 
Although the OCP might seem nonconvex using these decision variables, one can turn it into an SDP by redefining the decision variables and, if desired, reversing the transformation to recover the original variables from the solution.
As the number of neighbours of each subsystem increases, the dimensions of all constraints remain unchanged except for \eqref{sec3_mrpi1} which increases linearly and hence, the computation time is not significantly altered.
Similar to the OCP \eqref{sec2_dcn_OCP}, the online OCP \eqref{OCP} is solved at each time instant and the optimal control input $v_i(t)=u_i(0|t)+K_{o_i}T_i(t)(z_i(t)-x_i(0|t))$ is applied to the $i$-th subsystem.



\section{Feasibility and Stability}

In this section, we prove that the online OCP \eqref{OCP} is recursively feasible and the resulting closed-loop system is input-to-state stable. In the sequel, we use the superscript $(\cdot)^*$ to refer to the optimal solution. The proof is inspired from \cite{kouvaritakis2016model,limon2008mpc} and is divided into three main steps. First, we discuss the convergence of the nominal state $x^*(t)=[x_1^{*^\top}(t),\hdots,x_M^{*^\top}(t)]^\top$ to the artificial equilibrium $x_{e}^*(t)=[x_{e_1}^{*^\top}(t),\hdots,x_{e_M}^{*^\top}(t)]^\top$. Then, we discuss the convergence of the artificial equilibrium $x_{e}^*(t)$ to the target point $x_r=[x_{r_1}^{\top},\hdots,x_{r_M}^{\top}]^\top$. Finally, we discuss the stability of the error dynamics between the actual state $z^*(t)=[z_1^{*^\top}(t),\hdots,z_M^{*^\top}(t)]^\top$ and the nominal state $x^*(t)$.
We start with the first step in Lemma \ref{sec4_lm1}.
\begin{lem}
    \label{sec4_lm1}
    The nominal state trajectory converges asymptotically to the artificial equilibrium trajectory.
\end{lem}
The proof uses standard MPC arguments (\cite{kouvaritakis2016model}) by comparing the optimal cost at time $t$, the tail sequence cost at time $t$ and the optimal cost at time $t+1$.
Second, we discuss the convergence of $x_e^*(t)$ to the target point $x_r$. For this purpose, we define ($\tilde{x}_r$,$\tilde{u}_r$) to be the projection of ($x_r$,$u_r$) onto the set of equilibrium points contained in the set ($\xi \mathcal{X}_i(t)$,$\xi \mathcal{U}_i(t)$). We also define the matrix $\tilde{P}$ such that $\tilde{P}-(A+BK(t))^\top \tilde{P} (A+BK(t)) - Q - K^\top(t) R K(t) \geq 0$. Note that such $\tilde{P}$ exists since $K(t)$ asymptotically stabilizes the actual global dynamics as shown in Proposition \ref{prop3}. Finally, we define $E(c,s) = \{x:\|x-c\|^2_{\tilde{P}} \leq s\}$. For brevity, we omit the dependence on $t$ in the proof of Lemma \ref{lem2}. This proof is inspired from \cite{limon2008mpc}, however, we prove that $\lim_{t \rightarrow \infty} (x^*(t)-x_e^*(t))=0$ implies $\lim_{t \rightarrow \infty} (x_e^*(t)-x_r)=0$ instead of $x^*(t)-x_e^*(t)=0$ implying $ x_e^*(t)-x_r=0$.
    
    \begin{lem}
        \label{lem2}
        If $x^*(t)$ and $x_e^*(t)$ are such that $\lim_{t \rightarrow \infty} (x^*(t)-x_e^*(t)) = 0$, then $\lim_{t \rightarrow \infty} (x_e^*(t)-\tilde{x}_r) = 0$.
    \end{lem}
    
    \begin{pf}
        According to the OCP \eqref{OCP}, $(x_e^*,u_e^*) \in \operatorname{int}(\mathcal{X} \times \mathcal{U})$. 
        Hence, there always exists $E(x_e^*,\beta_1)$ such that $(x,K(x-x_e^*)+u_e^*) \in \gamma \mathcal{X} \times \gamma \mathcal{U}$ for all $x \in E(x_e^*,\beta_1)$ where $\gamma \in (0,1)$. 
        Note that $E(x_e^*,\beta_1)$ is invariant since it is defined using the level sets of a Lyapunov function. 
        We now define an equilibrium point $(\bar{x}_e,\bar{u}_e) \in \operatorname{int}(\mathcal{X} \times \mathcal{U})$ such that $\|\bar{x}_e-\tilde{x}_r\|_S^2 = \beta_2 \|{x}_e^*-\tilde{x}_r\|_S^2$ and $\|\bar{x}_e-x_e^*\|^2_{\tilde{P}}=(1-\lambda)\|\tilde{x}_r-x_e^*\|^2_{\tilde{P}}$ where $\beta_2 \in (0,1)$ and $\lambda \in (0,1)$ is chosen sufficiently large such that $(0,K(\bar{x}_e-x_e^*)+\bar{u}_e-u_e^*) \in (1-\gamma)\mathcal{X} \times (1-\gamma) \mathcal{U}$. 
        Next, we define $E(\bar{x}_e,\beta_3) \subset E(x^*_e,\beta_1)$ such that $x_e^* \in \operatorname{int}(E(\bar{x}_e,\beta_3))$. 
        This set is invariant since it is defined using the level sets of a Lyapunov function. Moreover, for all $x \in E(\bar{x}_e,\beta_3)$, $(x,K(x-\bar{x}_e)+\bar{u}_e)=(x,K(x-x_e^*)+u^*_e)+(0,K(\bar{x}_e-x_e^*)+\bar{u}_e-u_e^*) \in \left( \gamma \mathcal{X} \times \gamma \mathcal{U} \right) \oplus  \left((1-\gamma) \mathcal{X} \times (1-\gamma) \mathcal{U}\right) = \mathcal{X} \times \mathcal{U}$.
        
        We now define $\beta_4 > 0$ such that $S>\beta_4\tilde{P}$ and the set $\hat{\mathcal{X}} = \{x:\|x-x_e^*\|_{\tilde{P}}^2 \leq \beta_5 \|x_e^*-\tilde{x}_r\|_{\tilde{P}}^2 \ \& \ (x_e^*-\bar{x}_e)^\top \tilde{P} |x-x_e^*| \leq \beta_5 \|x_e^*-\tilde{x}_r\|_{\tilde{P}}^2 \}$ where $0<\beta_5<\beta_4(1-\beta_2)/3$. 
        Hence, $ \|x-\bar{x}_e\|_{\tilde{P}}^2 
        = 
        \|x-x^*_e\|_{\tilde{P}}^2 + 2 (x_e^*-\bar{x}_e)^\top \tilde{P} (x-x_e^*) + (1-\lambda)\|x_e^*-\tilde{x}_r\|_{\tilde{P}}^2 
        \leq 
        \|x_e^*-\tilde{x}_r\|_{3\beta_5\tilde{P}+(1-\lambda)\tilde{P}}^2$ for all $x \in E(x_e^*,\beta_6) \subset \hat{\mathcal{X}}$. 
        Note that $E(x_e^*,\beta_6)$ can be also chosen to lie inside $E(\bar{x}_e,\beta_3) \subset E(x_e^*,\beta_1)$ and hence is invariant. 
        Note also that $\|x-\bar{x}_e\|_{\tilde{P}}^2+\|\bar{x}_e-\tilde{x}_r\|_S^2 \leq \|x_e^*-\tilde{x}_r\|_S^2$ for all $x \in E(x_e^*,\beta_6)$ if $(1-\beta_2)S-(1-\lambda+3\beta_5) \tilde{P} \geq 0$ or equivalently, $\beta_4(1-\beta_2)-(1-\lambda+3\beta_5) \geq 0$. This can be achieved if $\lambda \geq 1-(\beta_4(1-\beta_2)-3\beta_5)$. Since $\beta_5<\beta_4(1-\beta_2)/3$, this condition on $\lambda$ does not contradict the aforementioned conditions on $\lambda$.
        
        We conclude the proof using contradiction. Assume that $\lim_{t \rightarrow \infty} (x^*(t)-x_e^*(t)) = 0$ and $\lim_{t \rightarrow \infty} (x_e^*(t) - \tilde{x}_r) \neq 0$. Hence, it is always possible to pick an arbitrarily large $t$ such that $x^* \in E(x_e^*,\beta_6) \subset (\bar{x}_e,\beta_3)$. In addition, there exists $\beta_7$ such that $\|x_e^*-\tilde{x}_r\| \geq \beta_7$ for infinitely many $t$. Since $E(x_e^*,\beta_6)$ and $E(\bar{x}_e,\beta_3)$ are both invariant and satisfy the state and input constraints, let $J^*$ and $\bar{J}$ be the cost corresponding to the optimal solution and the feasible solution aiming to converge to $\bar{x}_e$ and note that $J^* < \bar{J}$. Note also that $\bar{J} \leq \|x^*-\bar{x}_e\|_{\tilde{P}}^2+\|\bar{x}_e-\tilde{x}_r\|_S^2$ due to the Lyapunov inequality that $\tilde{P}$ satisfies. Since $x^* \in E(x_e^*,\beta_6)$, then $\|x^*-\bar{x}_e\|_{\tilde{P}}^2+\|\bar{x}_e-\tilde{x}_r\|_S^2 \leq \|x_e^*-\tilde{x}_r\|_S^2$. But, $\|x_e^*-\tilde{x}_r\|_S^2 \leq \|x_e^*-x_r\|_S^2 \leq J^*$ which contradicts the optimality of $J^*$.
        \hfill
        $\square$
    \end{pf}
    
    Finally, we discuss the input-to-state stability of the global closed-loop system under the MPC scheme \eqref{OCP}. Recall from \eqref{sec2_dyn_glb} that the global system is given by $z(t+1)=(A_d+E\tilde{A}_gC)z(t)+Bv(t)+d(t)$. Recall also that the optimal input is given by $v^*(t)=u^*(t)+K^*(t)(z^*(t)-x^*(t))$ where $K^*(t)=\operatorname{diag}(K_{o_1}T_1^*(t),\hdots,K_{o_M}T_M^*(t))$ and $u^*(t)=u^*(0|t)=[u_1^{*^\top}(0|t),\hdots,u_M^{*^\top}(0|t)]^\top$. Hence, the closed-loop system is given by $z^*(t+1)=A_dz^*(t)+Bu^*(t)+BK^*(t)(z^*(t)-x^*(t))+E\tilde{A}_gCz^*(t)+d(t)$.
    By defining $e^*(t)=z^*(t)-x^*(t)$ and $\eta(t)=d(t)-x^*(t+1)+Ax^*(t)+Bu^*(t)$, the closed-loop system becomes,
    \begin{equation}
        \label{sec4_sys}
            e^*(t+1) = (A+BK^*(t))e^*(t)+\eta(t).
    \end{equation}
    While $e^*(t)$ is the state of this system, $x^*(t)$ and $u^*(t)$, together with $d(t)$, are exogenous signals to this system. While $d(t)$ is bounded by assumption, $x^*(t)$ and $u^*(t)$ are bounded since they are the outputs of the proposed MPC problem (15) which consider compact constraint sets. Hence, we prove input-to-state stability with respect to these bounded exogenous signals. Note that $x^*(t+1)$ is not necessarily equal to $Ax^*(t)+Bu^*(t)$ where $A=\operatorname{diag}(A_1,\hdots,A_M)+EA_gC$. This is because $x^*(\cdot)$ and $u^*(\cdot)$ are the nominal states and inputs which satisfy $x^*(t+1)=\operatorname{diag}(A_1,\hdots,A_M)x^*(t)+Bu^*(t)$.   
    Moreover, the considered system is switched since $K^*(t)$ is adapted at each time instant. Hence, the closed-loop system under the proposed MPC scheme can be considered as an uncertain switched system with bounded disturbances.

    
    \begin{thm}
        The MPC scheme \eqref{OCP} is recursively feasible and the global closed-loop system \eqref{sec4_sys} under this scheme is input-to-state stable with respect to $\eta(t)$.
    \end{thm}
    
    \begin{pf}
        Recursive feasibility is proven using standard MPC arguments (\cite{kouvaritakis2016model}) by making use of the tail sequence. Regarding stability, we first consider the disturbance-free system $e^*(t+1)=(A+BK^*(t))e^*(t)$. The matrix $A+BK^*(t)$ is always Schur for any $K^*(t)$ since the constraints derived in Propositions \ref{sec3_prop2} and \ref{prop3} ensure that $P-(A+BK^*(t))^\top P (A+BK^*(t)) \geq \Gamma(t) - C^\top \tilde{A}_g C - C^\top L D(t) L C \geq \epsilon I_n$ at the $t$-th time instant for all $t$. Since this condition is always satisfied using the same matrix $P$, then $V(e^*)=\|e^*\|^2_P $ is a Lyapunov function for the disturbance-free system independent of $K^*(t)$.
        Now, we consider the closed-loop system \eqref{sec4_sys}. Using the Lyapunov function $V(e^*)$ and defining $\eta(t)=d(t)-x^*(t+1)+Ax^*(t)+Bu^*(t)$ , we can deduce that $V((A+BK^*(t))e^*(t)+\eta(t))-V(e^*) \leq -(\epsilon-\delta)\|e^*(t)\|^2 - \delta \|e^*(t)\|^2 + 2\|\eta(t)\|_P \|(A+BK^*(t))e^*(t)\|_P + \|\eta(t)\|_P^2$ where $0 < \delta < \epsilon$. To prove input-to-state stability, it suffices to ensure that $\delta \|e^*(t)\|_2^2 \geq 2\|\eta(t)\|_P\|(A+BK^*(t))e^*(t)\|_P + \|\eta(t)\|_P^2$ for some $\zeta > 0$ such that $\|e^*(t)\|\geq\zeta$. 
        We denote the maximum eigenvalue of $P$ by $\lambda_{max}(P)$ and the maximum norm of $\eta(t)$ by $\eta_{max}$. Since $\|(A+BK^*(t))e^*(t)\|_P < \|e^*(t)\|_P$ and $\|(\cdot)\|_P \leq \sqrt{\lambda_{max}(P)}\|(\cdot)\|$, it suffices instead to ensure that $\delta \|e^*(t)\|^2 - 2 \lambda_{max}(P)\eta_{max}\|e^*(t)\|-\lambda_{max}(P)\eta_{max}^2 \geq 0$. By computing the roots of the convex quadratic function on the left-hand side of the inequality, it is easy to verify that this inequality holds for $\|e^*(t)\| \geq \zeta = \left(\lambda_{max}(P)+\sqrt{\lambda_{max}^2(P)+\delta\lambda_{max}(P)}\right)\eta_{max}/\delta$. Hence, input-to-state stability is guaranteed.
        \hfill
        $\square$
    \end{pf}
    

\section{Numerical Results}

In this section, we investigate the efficacy of the proposed scheme by comparing it to existing schemes in the literature. In particular, we compare the proposed scheme (ADP+LRN) to robust MPC (ROB) (\cite{mayne2005robust}), adaptive MPC (ADP) (\cite{parsi2021distributed}) and learning-based MPC (LRN) (\cite{aswani2013provably}). Note that ADP is an extension of the adaptive MPC scheme in \cite{lorenzen2019robust} to the class of interconnected systems. We solve all optimization problems using MATLAB with YALMIP (\cite{lofberg2004yalmip}), Mosek and Gurobi on a computer equipped with a 1.9-GHz Intel core i7-8550U processor.

In the simulation example, we use a network of five double integrators connected in series as in \cite{parsi2021distributed}. 
We consider each double integrator as a subsystem with the dynamics in \eqref{sec2_dcn_dyn} where $A_{i} = \begin{bmatrix} 1 & 1 \\ 0 & 1 \end{bmatrix}$, $B_i=[0.5 \ 1]^\top$, $C_i=[0 \ 1]$ and $E_i=[0.05 \ 0.1]^\top$ for all $i \in \{1,2,3,4,5\}$. 
All states and inputs are constrained between $-2$ and $2$. 
The cost function matrices are $Q_i=I_2$, $R_i=0.1$ and $S_i=10I_2$ for all subsystems and the prediction horizon is $5$. 
The noise $d(t) \in \mathbb{R}^{2}$ is extracted uniformly with the first entry lying between $-0.025$ and $0.025$ and the second between $-0.05$ and $0.05$. 
The unknown true value of the parameters $a_{12}$ and $a_{54}$ is $1$, whereas that of $a_{21}$, $a_{23}$, $a_{32}$, $a_{34}$, $a_{43}$ and $a_{45}$ is $0.5$. The parameters $a_{12}$ and $a_{54}$ are assumed to lie initially between $0$ and $4$, whereas all other parameters are assumed to lie initially between $0$ and $2$.
The unknown true values of the parameters are not used within the proposed framework and, by design, communication is not used in the adaptation phase to locally solve the OCPs.
Each subsystem is required to follow a randomly generated piecewise constant reference changing every $100$ seconds and constrained between $-2$ and $2$. The simulation time is $1000$ seconds and the initial condition is $[0 \ 0]^\top$ for all subsystems.

\begin{figure}
    \centering
    \includegraphics[scale=0.25]{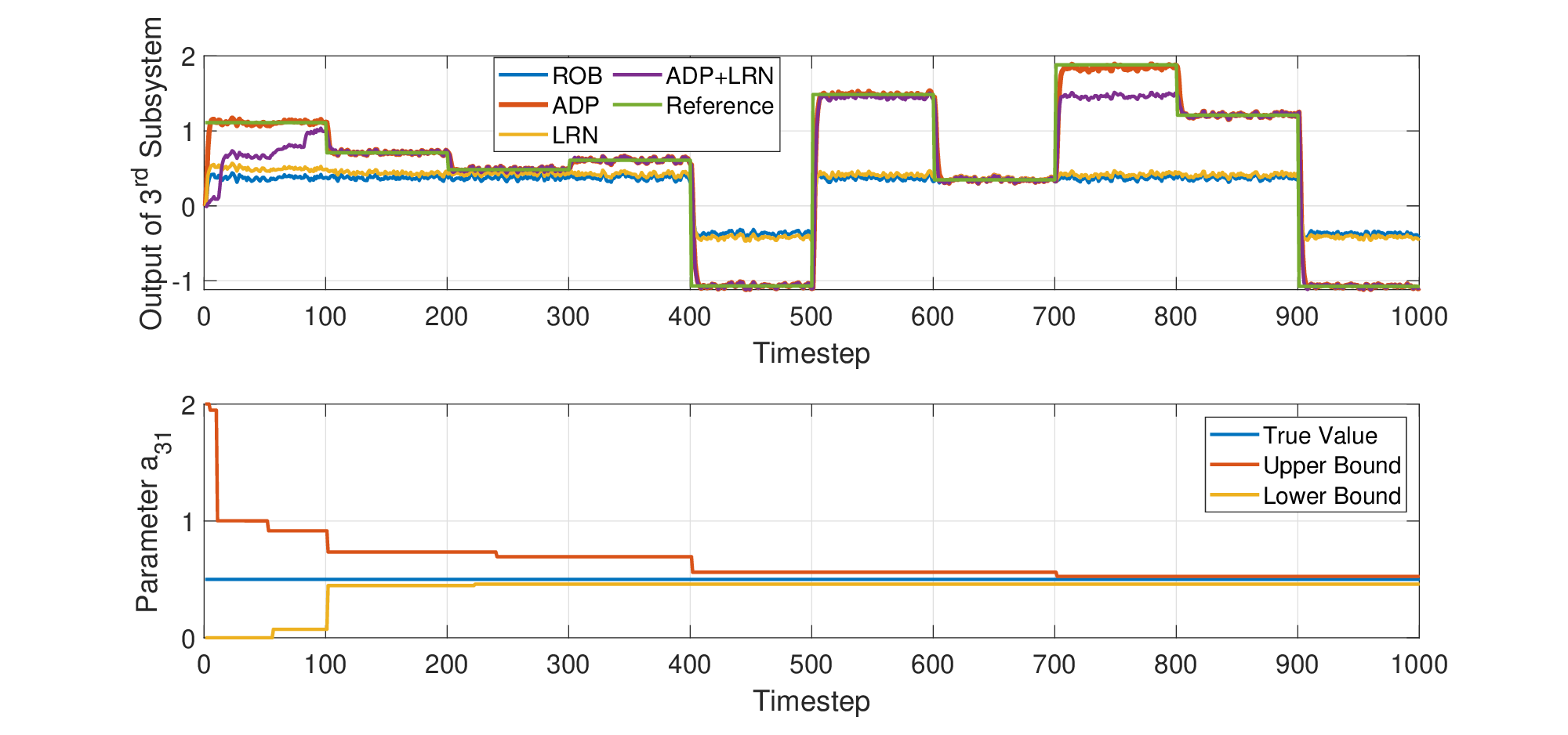}
    \caption{Top: Output trajectory of the third subsystem. Bottom: Bound estimates of the parameter $a_{31}$.}
    \label{fig:performance}
\end{figure}

Fig.\ref{fig:performance}(Top) shows the closed-loop performance of the third subsystem for each scheme. As shown in this figure, ROB and LRN generally fail to track the reference trajectory since LRN updates the model only while keeping the initial tube and tightened constraint sets unchanged and ROB updates nothing. On the contrary, ADP follows the reference successfully since it updates the tube and the tightened constraint sets online. Similarly, ADP+LRN generally follows the reference (due to the online shrinkage of the uncertain parameter sets as shown in Fig.\ref{fig:performance}(Bottom) for the parameter $a_{31}$) expect for extreme references close to the actual state constraint bounds due to the conservative constraint tightening.

\begin{figure}
    \centering
    \includegraphics[scale=0.25]{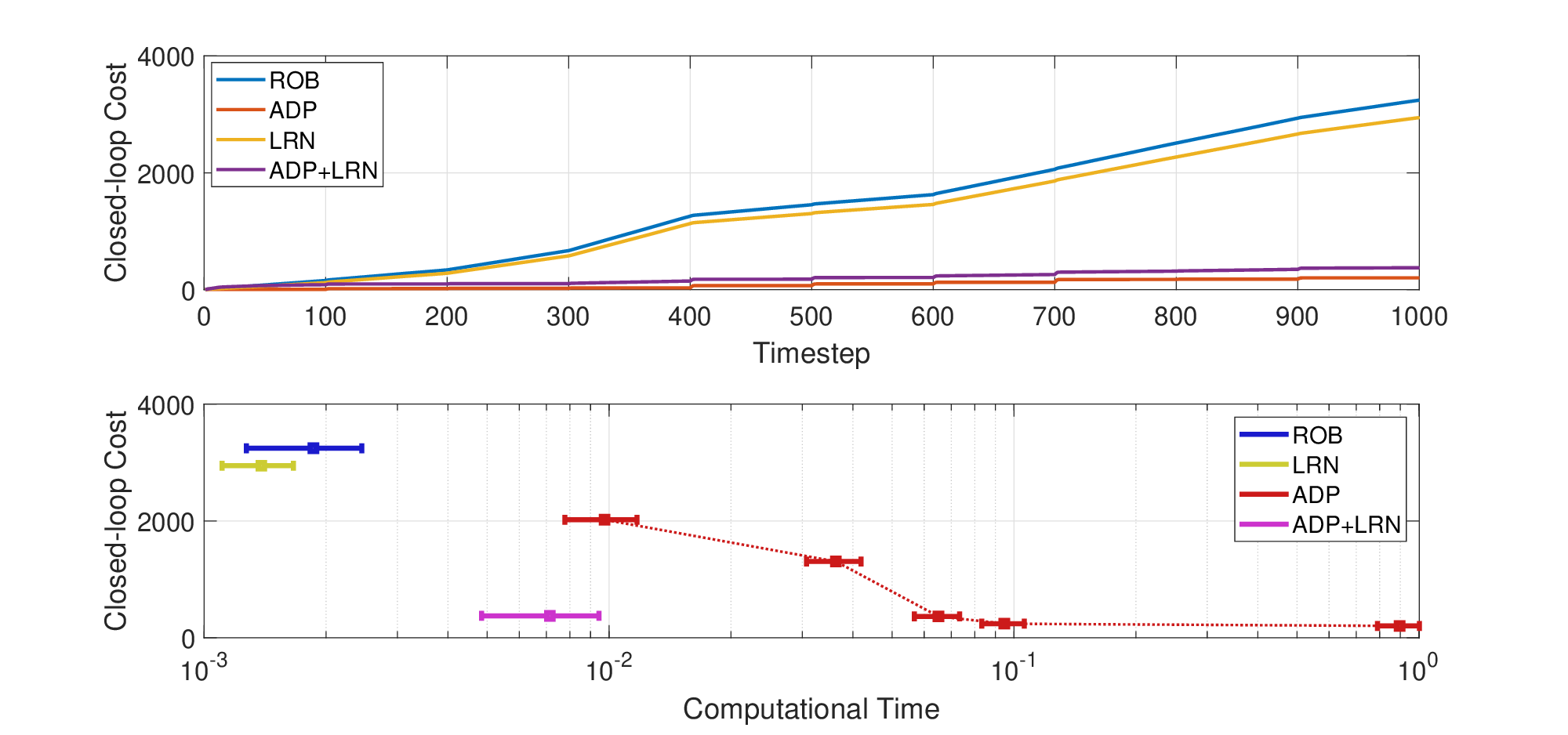}
    \caption{Top: closed-loop cost, Bottom: computation time mean (square) and standard deviation (bar).}
    \label{fig:comtime}
\end{figure}

The resulting behavior of each scheme has a direct effect on the closed-loop cost shown in Fig.\ref{fig:comtime}(Top). ROB has the highest cost, followed by LRN whose cost is slightly lower due to updating the model. The costs of ADP+LRN and ADP are considerably lower with ADP having the lowest cost. 
This is because ADP can quickly track the reference, even in extreme cases, since this scheme can directly handle parametric uncertainties unlike ADP+LRN which lumps all uncertainties in an additive term. 
ADP+LRN, however, results in a decentralized OCP which requires significantly reduced computations and communication among neighbours. On the other hand, ADP yields a distributed OCP that is solved using distributed optimization techniques (e.g. ADMM) leading to extensive communication and computations.

Fig.\ref{fig:comtime}(Bottom) shows the mean and standard deviation of the computation time required by each scheme (including different versions of ADP which differ in the maximum ADMM iterations allowed) and the corresponding closed-loop cost. We use the ADP scheme whose computation time is around 1 second in the other figures. It is found, however, that if we run ADMM for around 0.1 second only, we still get a relatively satisfactory performance. Running ADMM for less time (e.g. same order of magnitude of the time required by ADP+LRN) result in severe closed-loop performance deterioration.
On the other hand, all other schemes have lower computation time due to their decentralized nature which do not require using distributed optimization techniques such as ADMM. Since the OCP of ADP+LRN is a semidefinite program, ADP+LRN requires slightly higher computation compared to LRN and ROB whose OCPs are quadratic programs. 

\section{Conclusion}

We propose an adaptive learning-based MPC scheme for uncertain interconnected systems. 
The developed scheme is found to yield a better trade-off between computation complexity and closed-loop performance compared to existing schemes. 
Future work include applying the developed scheme to real-world applications.

\begin{ack}                               
The authors are grateful to Prof. Roy Smith, Prof. Andrea Iannelli, Dr. Goran Banjac and Dr. Anil Parsi for the fruitful discussions on the topic.
\end{ack}

{\small
\bibliographystyle{agsm} 
\bibliography{autosam}} 

\end{document}